\documentclass{article}
\usepackage{spconf,amsmath,graphicx}

\usepackage{url}
\usepackage{hyperref}

\newcommand{\dq}[1]{\href{https://stackoverflow.com/questions/#1/}{$Q_{#1}$}}

\title{StacerBot: A Stacktrace Search Engine for Stack Overflow}
\name{Md Abdullah Al Alamin}
\address{University Of Calgary\\Email: mdabdullahal.alamin@ucalgary.ca}

\begin{document}

\maketitle

\begin{abstract}
We as software developers or researchers very often get stacktrace error messages while we are trying to write some code or install some packages. Many times these error messages are very obscure and verbose; do not make much sense to us. There is a good chance that someone else has also faced similar issues probably shared similar stacktrace in various online developers' forums. However traditional google searches or other search engines are not very helpful to find web pages with similar stacktraces. In order to address this problem, we have developed a web interface; a better search engine: as an outcome of this research project where users can find appropriate stack overflow posts by submitting the whole stacktrace error message. The current developed solution can serve real-time parallel user queries with top-matched stack overflow posts within 50 seconds using a server with 300GB RAM. This study provides a comprehensive overview of the NLP techniques used in this study and an extensive overview of the research pipeline. This comprehensive result, limitations, and computational overhead mentioned in this study can be used by future researchers and software developers to build a better solution for this same problem or similar large-scale text matching-related tasks.
\end{abstract}

\section{Introduction} \label{sec:intro}

\begin{figure}[t]
\centering
\includegraphics[scale=0.25]{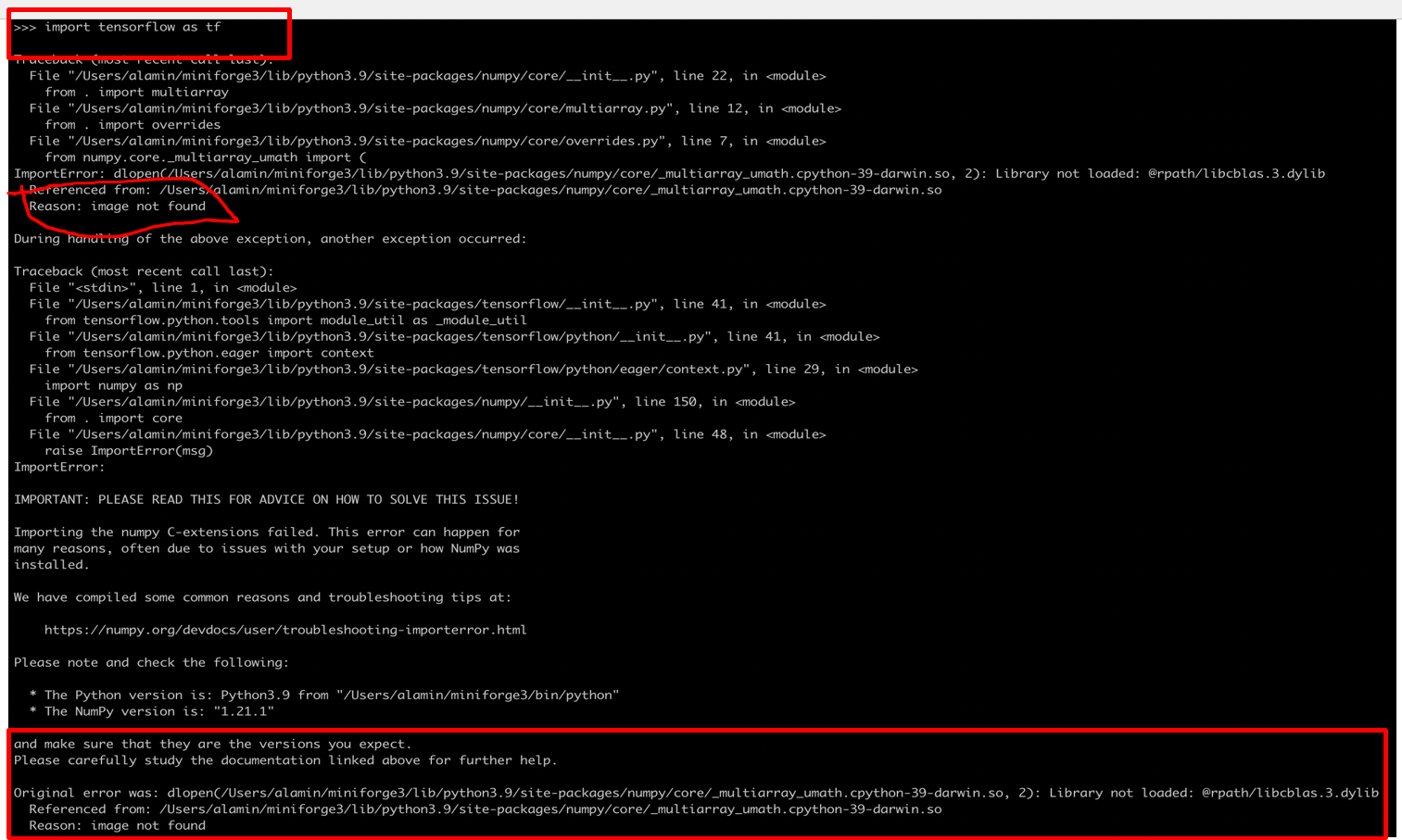}
\caption{Another example of a long stacktrace error message}
\label{fig:stacktrace_difficult}
\end{figure}

\textbf{Problem background.} We often get error messages  when we try to develop some application, install some packages, set up a development environment. We get these error messages called stacktrace (Fig. \ref{fig:stacktrace}). So, most of the time we select a few important lines from the error messages and try to make a google search or Stack Overflow search to find relevant Stack Overflow posts to solve the issue. Most of the time this approach works. However many times this does not work specially in a situation where the stacktrace is verbose with generic information (Fig. \ref{fig:stacktrace_difficult}). In these situations usually, even through there are Stack Overflow posts with a similar stacktrace a google search does not provide Stack Overflow relevant Posts link because that search engine is not designed to search and find stacktrace error message. For example, for the above error message in Fig. \ref{fig:stacktrace_difficult} there is an Stack Overflow post ``TensorFlow 2.5 Mac M1 - Installing problem compatibility with NumPy library / Conda env'' in \dq{68996176} which has an accepted solution 
\begin{verbatim}
conda install -c conda-forge openblas
\end{verbatim}
to solve the issue. Interestingly this solves the problem but a google search with sub-string of the stacktrace fails to get the link of this stack overflow post and its solution.

\textbf{Problem Motivation.}
Nowadays, there are many online forums --Stack Overflow, GitHub Issue page-- where developers share their error messages and ask for help from the community members. Many times the community members provide useful suggestions on how to overcome those issues. So, such discussions are of great value to the users who are facing similar problems. However, the standard search engines -- Google search or Stack Overflow search engine -- are not designed to search and match the whole stacktrace error message which can be hundreds of lines in some cases. So, the standard search engines fail to search march similar stacktrace error messages. In this research problem, we intend to address this issue.

\textbf{Research Objective.}
There are some other studies that try to find a solution from analyzing stacktrace error messages\cite{arnold2007stack, wong2014boosting, gu2019does}. However, the research objective of this study is to \textit{explore the computational limitations of different NLP techniques to build a \textbf{search engine} to address this issue.} There are many NLP techniques that can be used to accomplish this task and for this project we employ a well known NLP technique called Term Frequency Inverse Document Frequency (TF-IDF)~\cite{ramos2003using} to build a document index and build a web interface which will provide top matched stack overflow posts based on the provided raw large stacktrace error message.

\textbf{Final Outcome.} The final outcome of this study is the comprehensive research methodology used in this study (Section \ref{sec:methodology}) as well as discussion (Section \ref{sec:discussion}) of some of the fundamental strengths and weaknesses of this study. See the Section \ref{sec:result} for details.

\textbf{Replication Package}: The code and data are shared in: \tiny
\url{https://github.com/al-alamin/SO-Error-Finder/}
\small

\begin{figure}[t]
\centering
\includegraphics[scale=0.30]{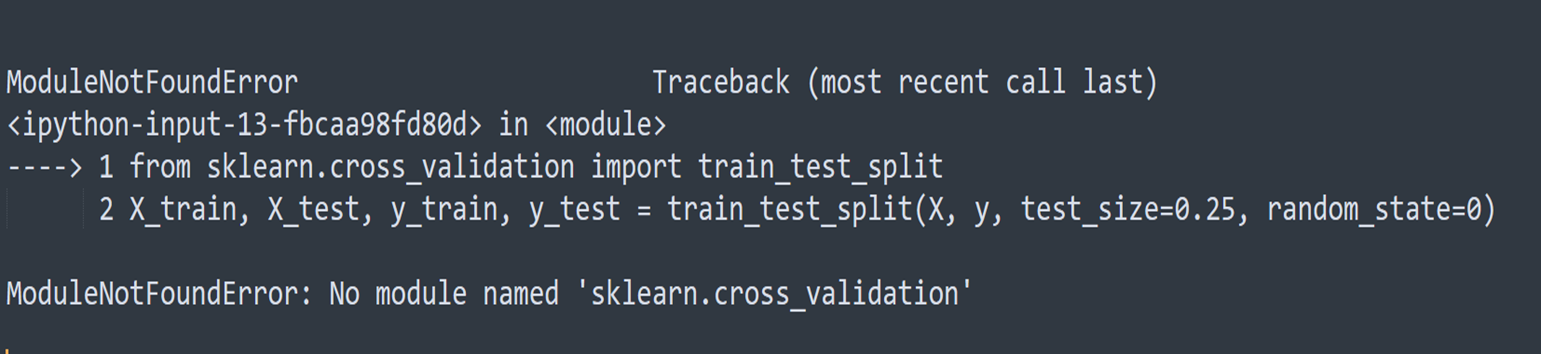}
\caption{Example of a stacktrace error message}
\label{fig:stacktrace}
\end{figure}

\begin{figure}[t]
\centering
\includegraphics[scale=0.22]{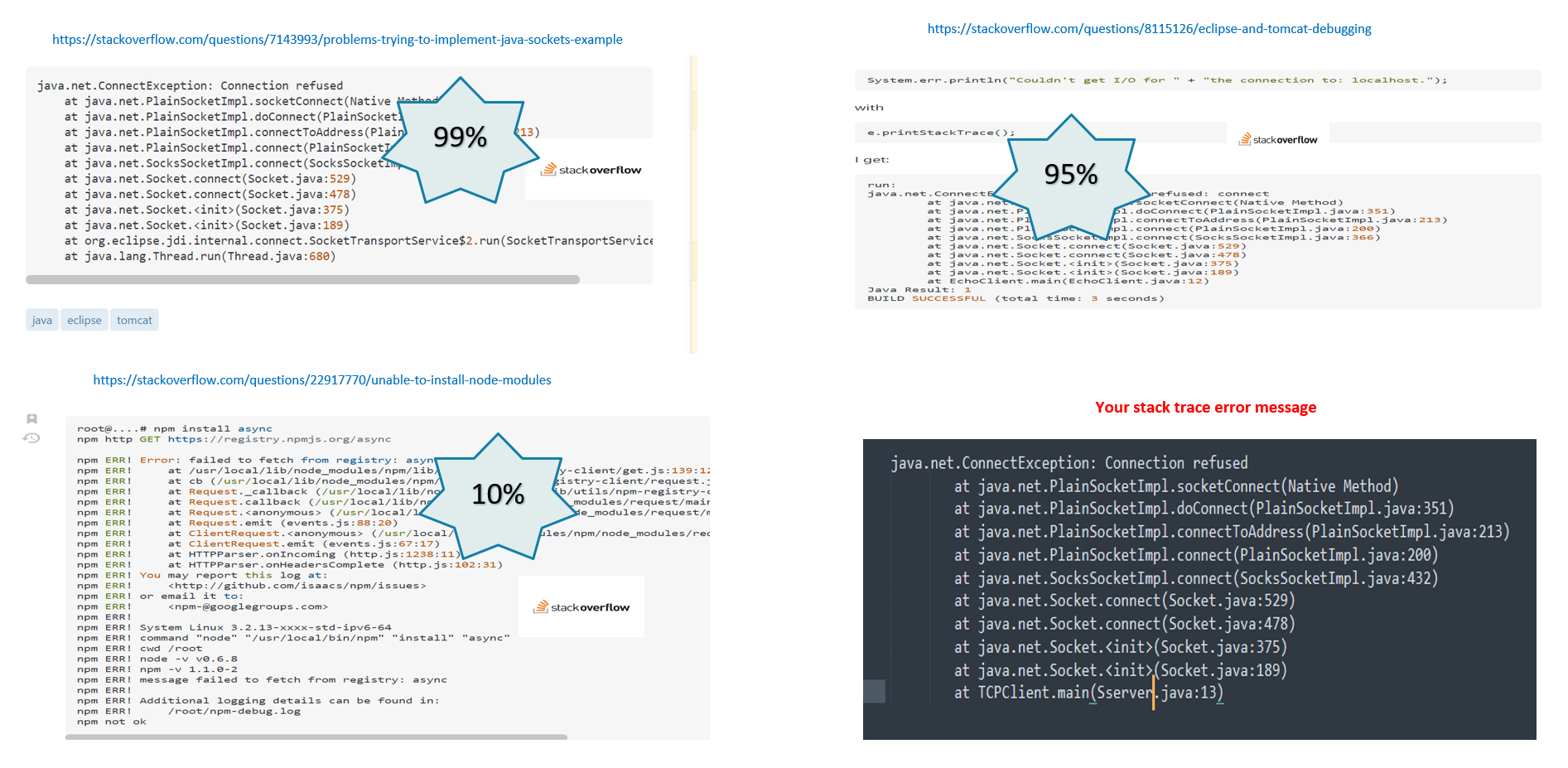}
\caption{Finding the appropriate Stack Overflow Posts with relevant stacktrace.}
\label{fig:similar_stacktrace}
\end{figure}
\section{Dataset} \label{sec:dataset}

For this study, we used the most popular Q\&A site, Stack Overflow (SO), where developers from diverse background discussion about various software and hardware related issues~\cite{barua2014developers}. For this study, We downloaded SO data dump~\cite{SOdump} of July 2021 which was one of the latest dataset available during the starting of this project. We used the contents of ``Post.xml'' file, which contained information about each post like the post's unique ID, type (Question or Answer), title, body, associated tags, creation date, view-count, etc. Our data dump included discussion of 12 years from July 2008 to July 2021 and contained around 50M posts. Total 11M users from all over the world participated in the discussions.

\textbf{Overview of dataset attributes.} ``Posts.xml'' contains 22 attributes:  ``Id", ``PostTypeId", ``AcceptedAnswerId", ``ParentId", ``CreationDate", ``DeletionDate", ``Score", ``ViewCount", ``Body", ``OwnerUserId", ``OwnerDisplayName", ``LastEditorUserId", ``LastEditorDisplayName", ``LastEditDate", ``LastActivityDate", ``Title", ``Tags", ``AnswerCount", ``CommentCount", ``FavoriteCount", ``ClosedDate", ``CommunityOwnedDate", ``ContentLicense". 

Among these 22 attributes we are mostly interested in the ``ID'': Question Id, ``Title'': Title of the post, ``Body'': contains the text of the stack overflow posts.

\begin{figure}[t]
\centering
\includegraphics[scale=0.3]{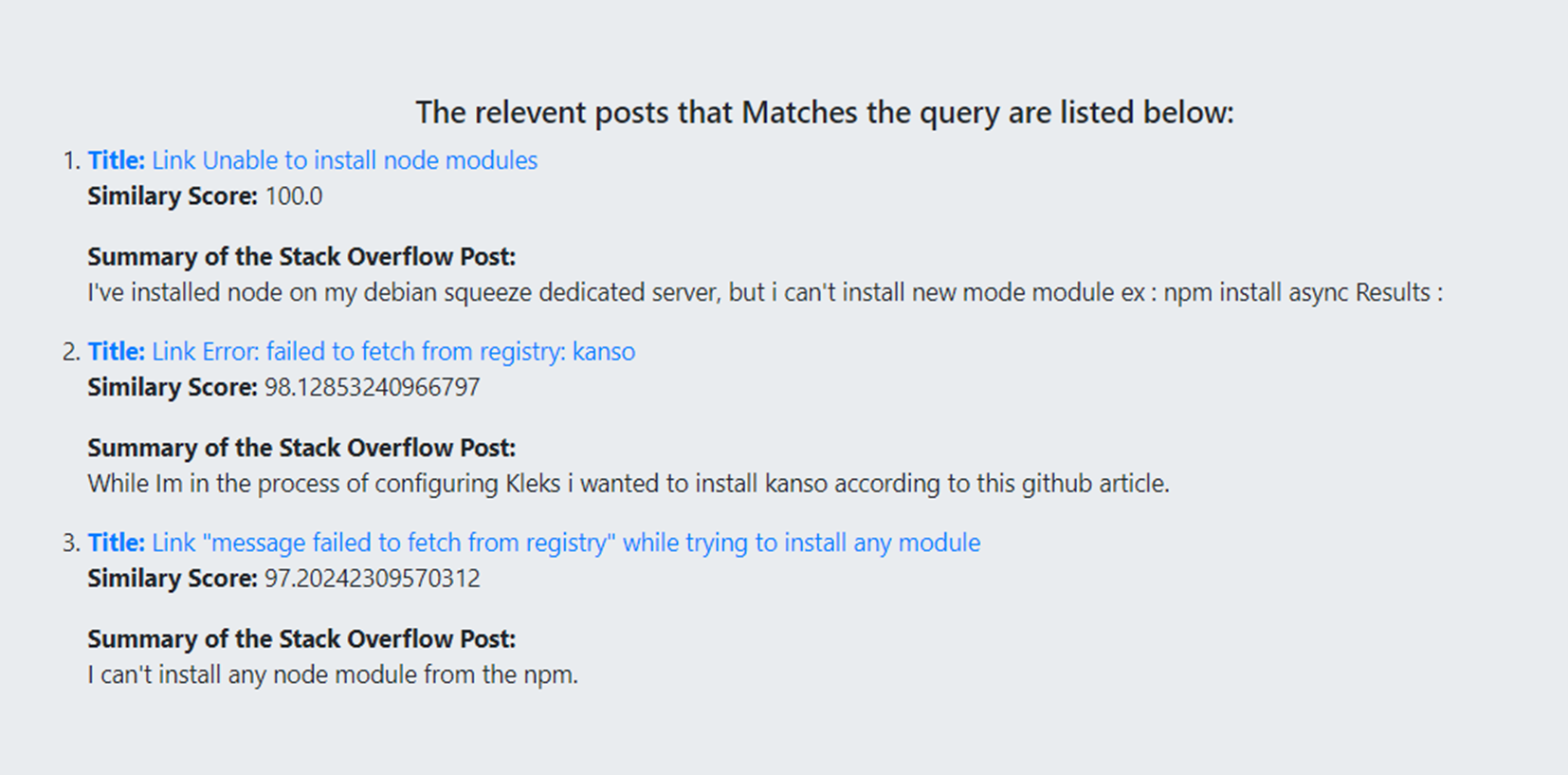}
\caption{Final output of our systems based the stacktrace query}
\label{fig:final_output}
\end{figure}

\begin{figure*}[t]
\centering
\includegraphics[scale=0.45]{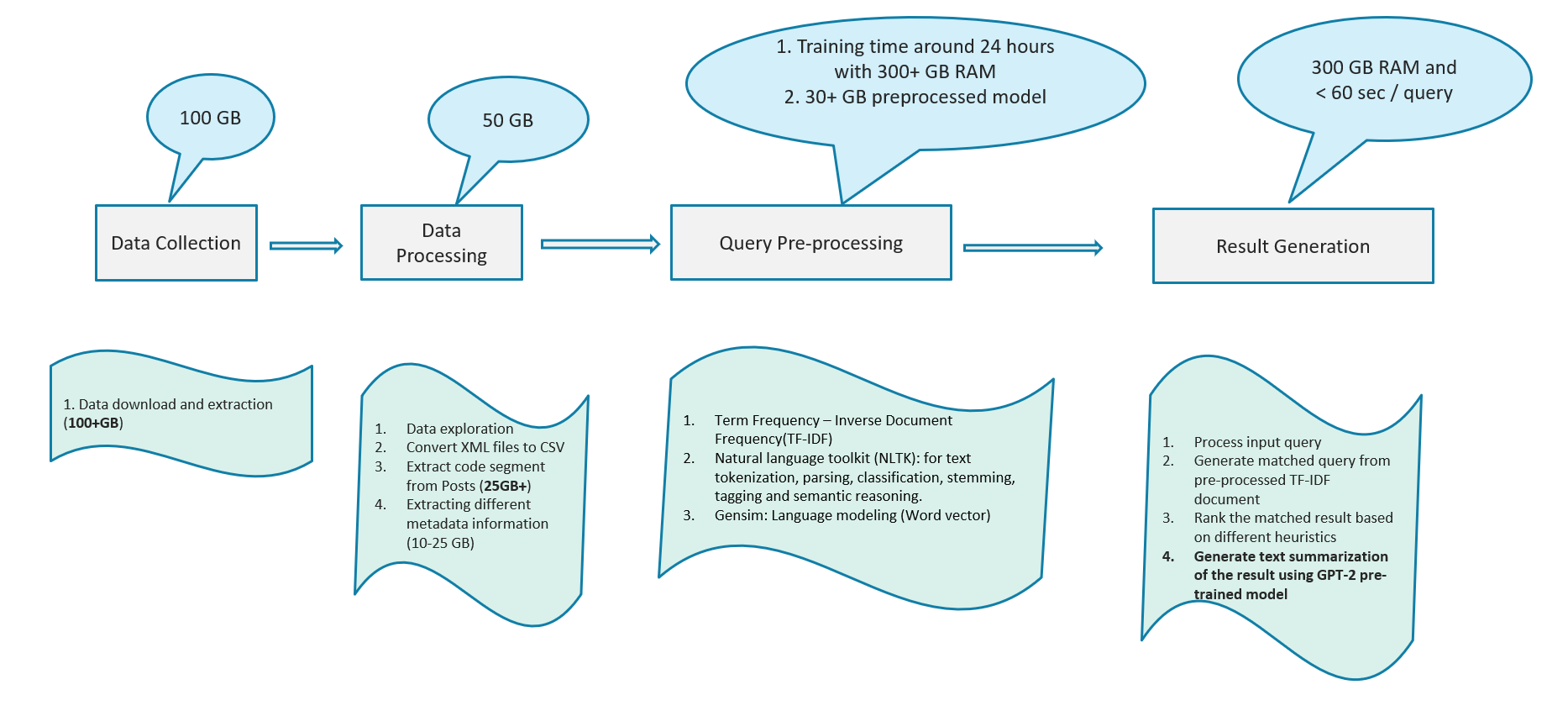}
\caption{Overall pipeline of research Methodology.}
\label{fig:stacktrace}
\end{figure*}

\begin{figure}[t]
\centering
\includegraphics[scale=0.28]{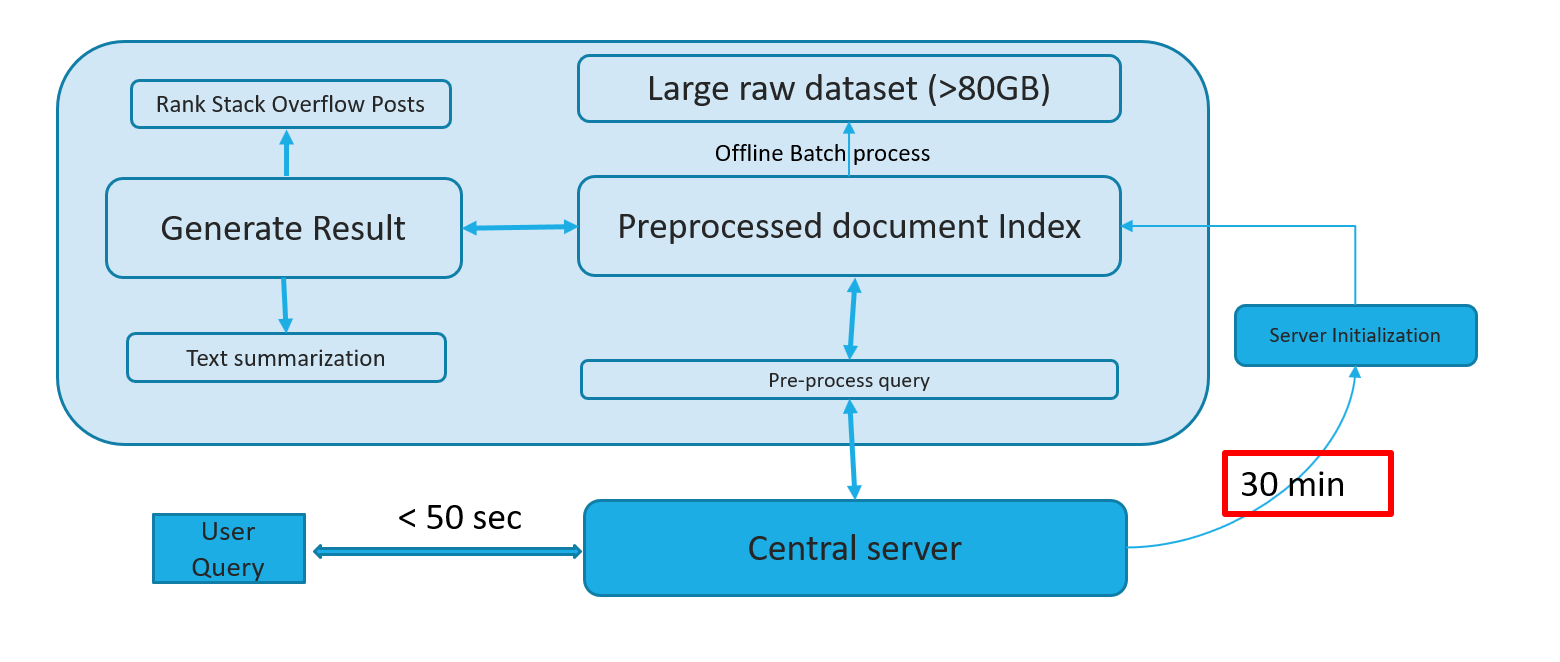}
\caption{Web server architecture of request and response}
\label{fig:webserver_architecture}
\end{figure}

\section{Experiment Setup} \label{sec:methodology}

In this Section the overall research methodology is described. We also provide detail description of the NLP technique used in this research on our selected dataset (\ref{sec:dataset})

\subsection{Data Collection}
First we collected Stack exchange data dump\cite{SOdump} got ``stackoverflow.com-Posts.7z'' file. Then we moved this dataset to ARC cluster where we extracted this .7z file and got Posts.xml file of size around 50GB. Later part of the experiments processed this XML file to a more usable format.

\subsection{Data processing}
After data collection, some of the data processing steps are described below.

\textbf{Converting XML files to CSV files.}
In the first step, first, we need to convert the XML file to a CSV file so that it can be extracted and processed easily via various NLP data processing libraries such as Pandas\cite{mckinney2011pandas}. During this project, I had to attempt several times to successfully convert this ``Posts.xml'' file to ``Posts.csv'' file because there was some syntax error in the original XML file that I needed to handle and at the same time I had to process this XML file sequentially because the popular LXML\cite{behnel2005lxml} library that I used can not process such a big XML file all at once. The final size of ``Posts.csv'' file was around 85GB.

\textbf{Extracting code segment from posts.} The second step of the task is to extract the code segments from the stack overflow post. Each Stack Overflow's post contains the code segments, i.e., the error messages of stacktrace inside a HTML tag called \textless code\textgreater Stacktrace code segments \textless /code\textgreater. We extract these code segments from stack overflow posts. In the later steps, we'll create a document Index of this code segment so that during query we can very easily find appropriate Stack Overflow posts. From our dataset, there were around \textbf{33M} Stack Overflow Posts that contained some code segment. We call this 33M extracted code segment our \textbf{Corpus $C$}. Corpus is an NLP jargon meaning a collection of text. The final size of our Corpus $C$ is around \textbf{25GB}.

\textbf{Extracting metadata information from posts}. The web interface that I demonstrated provides the relevant Stack Overflow posts with some useful information. To this end we also extracted: ``Id'', ``PostTypeId'', ``AcceptedAnswerId", ``ParentId", ``CreationDate", ``ViewCount", ``Score". All of these metadata were used by the heuristic ranking algorithm to ranking the matched posts.

\subsection{Query pre-processing.}
For this project, we need to create a document index matrix of the 33M code segments that we have extracted. In this way, when a new stacktrace is provided by the user we can very fast match with these 33M code segments and get an array of similarity matches of 33M entries. When the Document Index is prepared this matching happens very fast, usually within 10 to 20 seconds. One of the limitations is that in order to match fast all the data is needed to be stored in the RAM and thus it takes around 300GB of RAM for this project.

For this project, we used Natural language toolkit (NLTK) NLTK~\cite{loper2002nltk} which is one of the most popular Python libraries for Natural Language Processing (NLP) tasks. It has a big community and in this project, it is used for various text processing tasks such as tokenization, parsing, stemming, tagging, etc. We also used the popular Gensim~\cite{vrehuuvrek2011gensim} python library for creating a bag of words and Document Index. In order to create the document index, there are some pre-requisite steps such as word tokenization, creating a word dictionary which is described below.

\subsubsection{NLP text pre-processing}
Here the widely used NLP techniques such as creating a word dictionary, bag of words are described in detail with examples.

\textbf{Word tokenization.}
This is a process to split a sentence into a list of words. First, we lowercase the whole stack trace and then tokenize it. For examples, ``I am a Master's students'' will be split in to an array of words [``i'', ``am'', ``Master's'', ``student''].

\textbf{Word Dictionary\cite{silberztein2004architecture}}. After word tokenization, we need to create a word dictionary. In this step, each unique word is assigned a unique integer value. This is a common NLP technique for faster processing. The dictionary of our final Corpus $C$ contains around \textbf{2M unique words} this is quite high compared to other NLP tasks. We have this big Dictionary because in general for NLP tasks during this process usually use different techniques such as removal of stop words~\cite{wilbur1992automatic} and for unknown English words they simply assign a specific int value. But for this project, we could not use such techniques because many of the stacktrace will contain unknown English words and which is very important for us to find a good match.

\textbf{Bag of words~\cite{zhang2010understanding}}. The next step is to create a bag of words which is basically a map containing the unique word id and its frequency, i.e., the total number of times it appears in each of the samples posts in our Corpus of Stack Overflow posts.

\textbf{Term Frequency Inverse Document Frequency (TF-IDF)~\cite{ramos2003using}}. This is basically a bag of word models of our text corpus. One additional benefit of using this is it considers the frequencies and weights it down. For example, if a word like ``function'' appears too many times in our text corpus then it would assign fewer weights to it compared to a word that is used less frequently (e.g., ``Http 403''). This is a quite useful feature in our case because it helps us to search and find posts that match most to the search query.

\textbf{Gensim Document Index~\cite{gensim_sim}}. The next step is to build the document index from our Corpus. Once this index is built we can use this to make efficient queries like ``How similar are each of the documents (i.e., stacktrace) in the index is to the query document''. For this project building, this Document Index of our Corpus of size 33GB requires around \textbf{24 hours with 250 GB of RAM}. Once this index is built the next step is to generate results for query stacktrace.

\subsubsection{Result Generation}
Using the document index, we can find Stack Overflow posts with percentage of the match with the query text. The following section describes how the results are generated, ranked, and presented.

\textbf{Matching using Document Index}. Let's provide an example of how stacktrace queires are matched with our Corpus. First, we tokenize the query text, then we create a word dictionary based on our previously built Dictionary and then we create a tf-idf query of the query text. Based on this we query the document index which returns an array of 33M entries each entry contains a number from 0 to 1 indicating the match with that query text and the Corpus text. For this project, we extracted the top 30 Stack overflow Posts that match most to your query document.

\textbf{Posts Ranking.} In the next step we employ some heuristics to rank the stack overflow posts. For example, if two Stack Overflow posts have the same similarity with the query text then the stack overflow post which has accepted answers will get a higher position in the rank. Similarly, in our dataset, each stack overflow question and answer is considered a stack overflow post. So if the question contains some of the query text as well as if some answers of that discussion also contains similar stacktrace then we assumed that this discussion may be more relevant to the user and thus provided a higher position in the final ranking. We used the metadata information to get all the additional information -- does it have accepted answers-- to rank the posts. In summary, the ranking techniques used in this study are heuristic and can be improved a lot by adding more rules. Finally after ranking the top three matched stack overflow posts are returned for display.

\textbf{Text Summarization}. So, from the ranking of the post, we get the top three stack overflow post. Now before presenting the results to the users we format it a little bit more (See Fig \ref{fig:final_output}). So from the extracted metadata, we collect the title of the stack overflow posts make it a clickable link so that the user can click the click and redirect to the original stack overflow post. Then we present the similarity score. After that from the metadata, we collect the text that was used in the original stack overflow post to describe the problem. Then for this project, I used a pre-trained GPT-2~\cite{radford2019language} model for text summarization. GPT-2 is a very well-known pre-trained NLP model for such tasks which was trained on 40GB of Internet text. We used this pre-trained model to generate a text summary which is 25\% of the original text. Our motivation to summarize was that the title of the question, similarity score as well as summary will help the users to choose which question might be more relevant to them. In this project, we observed it takes around \textbf{5 sec} in CPU and \textbf{0.3 sec} in GPU for the model to summarize the three posts. Again, the output format used in this project can be improved a lot for real-world deployment.

\section{Result} \label{sec:result}
In this Section, we discuss the final output of this project.

\subsection{Web interface: real-time query processing.}\label{web-interface}
One of the outcomes of this project is to build a tool that actually returns top-matched stack overflow posts given a stacktrace error message. To this end, we build a web interface using Flask Python web framework~\cite{grinberg2018flask}. Figure \ref{fig:webserver_architecture} provides a high level architecture of the web interface. For example, there is a central server that first initializes, i.e., loads up the document index of our Corpus. This takes around \textbf{30 min and 300GB of RAM}. After that, the webserver is ready to process any query in real-time -- 50 sec-- and as this uses HTTP protocol and every module is independent parallel requests can also be processed. First, the webserver pre-processes in the query text as mentioned in the previous subsection. After that, it queries the document index and finds the top 30 matched stack overflow questions which take around 20-35 sec. When the top questions are prepared then heuristics ranking algorithm, text summarization, result formatting is done using metadata information. This whole process takes around \textbf{40-45 sec} and the user is provided an HTTP response similar as Fig \ref{fig:final_output}.

\textbf{Quality of the result.} During and after developing the systems we found the systems perform quite well in providing relevant Stack Overflow questions. For example, if take the following code segments from the Stack Overflow posts \dq{9626990} and use our system to make a prediction it provides three stack overflow posts. ``Error: failed to fetch from registry: kanso'' in \dq{26136411} with 98\% match and `message failed to fetch from registry while trying to install any module'' in \dq{12913141} with 97\% match. Both of which contain similar stacktrace with npm node error messages. The video demonstration provides better visualization of the output.

\tiny{}
\begin{verbatim}
npm ERR! Error: SSL Error: SELF_SIGNED_CERT_IN_CHAIN
npm ERR!     at ClientRequest.<anonymous> (/usr/lib/node_modules/npm/node_modules/request/main.js:252:28)
npm ERR!     at ClientRequest.emit (events.js:67:17)
npm ERR!     at HTTPParser.onIncoming (http.js:1261:11)
npm ERR!     at HTTPParser.onHeadersComplete (http.js:102:31)
npm ERR!     at CleartextStream.ondata (http.js:1150:24)
npm ERR!     at CleartextStream._push (tls.js:375:27)
npm ERR!     at SecurePair.cycle (tls.js:734:20)
npm ERR!     at EncryptedStream.write (tls.js:130:13)
npm ERR!     at Socket.ondata (stream.js:38:26)
npm ERR!     at Socket.emit (events.js:67:17)
npm ERR! Report this *entire* log at:
npm ERR!     <http://github.com/isaacs/npm/issues>
npm ERR! or email it to:
npm ERR!     <npm-@googlegroups.com>
npm ERR! 
npm ERR! System Linux 2.6.38-13-generic
npm ERR! command "node" "/usr/bin/npm" "install" "jed"
npm ERR! node -v v0.6.12
npm ERR! npm -v 1.0.104
\end{verbatim}
\small

\subsection{Outcome}
The outcome of the study is two fold. The first one is a theoretical discussion of the NLP technique that is used in this study. Future works can benefit from this to solve similar problems. Then the second part is the discussion of the computational complexity. This may provide valuable insights for future researchers who want to undertake similar tasks. For example, the very excessive \textbf{exponential memory requirement} of the current model may be a good indication to look for other approaches depending on the system requirement and limitations.

As we mentioned earlier the web interface requires around 300GB of RAM and thus it is not possible for us to host it on any free server. However we provide a Video demonstration of the web-interface: \tiny{\url{https://github.com/al-alamin/SO-Error-Finder/blob/master/Project/Project_demonstration.mov}}
\small

\begin{figure}[t]
\centering
\includegraphics[scale=0.220]{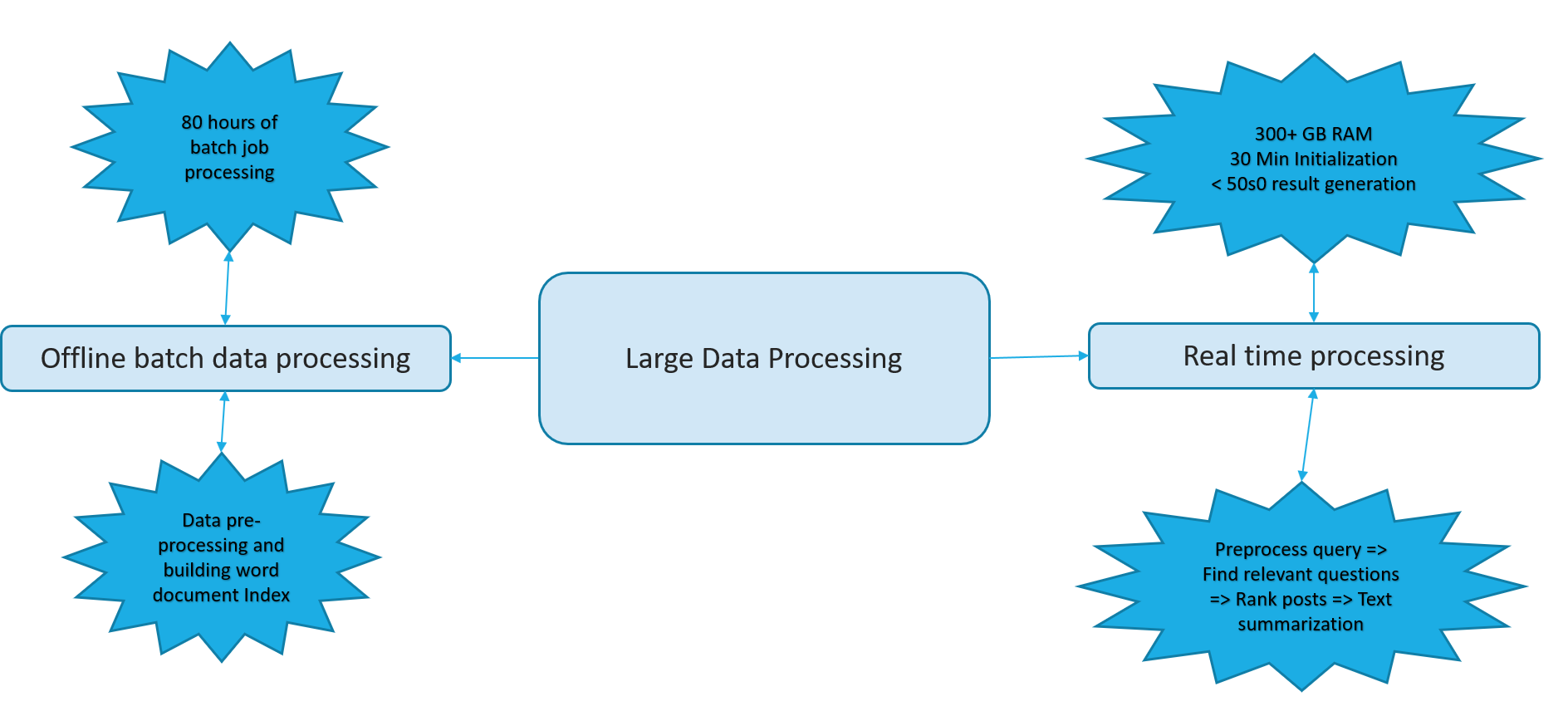}
\caption{Overview of asynchronous and real-time computational complexity }
\label{fig:computational_overview}
\end{figure}


\section{Discussion} \label{sec:discussion}
In this Section, we provide a summary of the computational challenges and discuss some of the limitations and future improvement scope for this project.

\subsection{Computational Challenges}
In this Section, we summarize the computational tasks to execute this project in both in terms of batch jobs and real-time jobs.

\subsubsection{Offline batch processing.} Offline batch jobs denote the tasks we needed to do beforehand. For example, converting the XML file to a CSV file, creating the document index for future faster search, extracting metadata information for faster result generation. In summation, all of these batch jobs took around 80 hours of time in total. Most of these offline jobs needed moderate RAM to expect the job for creating document index which needed around 300GB RAM.

\subsubsection{Real-time processing.} Real-time jobs denote the tasks of the web interface (Section \ref{web-interface}) where users query is processed by the web interface and this requires around 300GB of RAM. If GPU can be used then the text summarized task reduces to 1\%. This web interface can generate results within 50 sec of the query.


\subsection{Limitations}
Some limitations of this study are discussed below:
\begin{itemize}
    \item It takes around 50 sec for the web interface to generate the result. In order to get useful, the results should be generated within 2 sec. There is no way the current architecture can do this. The whole monolithic design will have to be broken up and designed in a more micro-service architecture.
    \item Many of the offline batch job tasks are written in Python library which uses only one CPU. Thus some of the tasks such as XML to CSV conversion code could have been written in a parallel way. However, considering the time improvement and the time it would take to rewrite those codes for parallelization; we skipped this part for this study.
    \item The solution used for this project has \textbf{exponential memory complexity} which is not ideal for large datasets. If the dataset size were twice as much as this then probably this system would have taken terabytes of RAM which is not always a feasible solution.
    \item There are some other NLP approaches to find similarity among multiple texts such as Cosine Similarity~\cite{faruqui2016problems} which were not explored properly during this study.
\end{itemize}

\subsection{Scope for improvement}
Some future improvement strategies are discussed below:
\begin{itemize}
    \item The heuristic algorithms used in this study to rank the posts can be improved. For example, posts score and view count can be considered in the ranking.
    \item The current systems only search for Stack Overflow. However, the dataset contains information related to other stack exchange sites such as ``AskUbuntu'', ``Serverfault'', ``Super User'' which also contains many of the relevant stacktrace. Future improvement can incorporate those.
\end{itemize}

\subsection{Experiment Environment Acknowledgement.} All the experiments for this study were conducted in the ARC and TALC cluster of the University of Calgary.
\section{Conclusion} \label{sec:conclusion}
For this project, we developed a computationally expensive web interface that allows users to search for relevant Stack Overflow Posts based on the stacktrace error messages. The main limitation for this project was exponential memory limitation which requires around 300 GB of RAM to build a document index and generate results in real-time. Despite some of the obvious limitations of this project, we do believe this system can be helpful for users to find useful relevant stack overflow posts from verbose, obscure stacktrace messages. The research methodologies and described strengths and limitations of the proposed approach will help future researchers or developers to make better decision for similar tasks.

\bibliographystyle{abbrv}
\bibliography{biblography}
\end{document}